\documentclass[aps,prl,showpacs,noshowkeys,amsmath,amssymb,amsfonts,reprint]{revtex4-1}
\usepackage{graphicx}
\usepackage{dcolumn}
\usepackage{bm}
\begin{document}
\title{Geometric Frustration of Colloidal Dimers on a Honeycomb Magnetic Lattice}
\author{Pietro Tierno}
\email{ptierno@ub.edu}                                          
\affiliation{
Estructura i Constituents de la Mat\`eria Universitat de Barcelona, 08028, Barcelona, Spain\\
Institut de Nanoci\`encia i Nanotecnologia,
Universitat de Barcelona, 08028, Barcelona, Spain}
\date{\today}
\begin{abstract}
We study the
phase behaviour and the collective
dynamics of interacting
paramagnetic colloids
assembled above a
honeycomb lattice of triangular shaped
magnetic minima.
A frustrated colloidal
molecular crystal is
realized when filling these
potential minima with exactly
two particles per pinning site.
External in-plane rotating fields
are used to
anneal the system
into different phases,
including long range ordered
stripes,  random fully packed loops,
labyrinth and disordered states.
At higher amplitude of the
annealing field, the dimer
lattice displays a two step
melting transition where the
initially immobile dimers
perform first localized
rotations and later break up by exchanging
particles
across consecutive lattice minima.
\end{abstract}
\maketitle
Geometric frustration
arises when the spatial arrangement of the
system elements
prevents simultaneous minimization of all
interaction energies,
and features at low temperature
a highly degenerate ground state~\cite{Moe06}.
Effects of such
phenomenon
manifest in disparate systems,
from classical
magnets~\cite{Wan50,New53},
to active matter~\cite{Shi09},
coupled lasers~\cite{Nix13},
complex networks~\cite{Dor08}
and quantum many-body systems~\cite{Gia11,Yi12,Seo13}.
Recent experiments with
size-tuneable microgel
particles~\cite{Han08}
have shown that
strongly confined colloids
represent a versatile model to investigate
geometrically frustrated states.
In contrast to lattices of interacting
nanoscale islands
such as artificial spin ice~\cite{Wan06,Nis13},
colloids feature
time and length scales
which are accessible via simple light microscopy,
combined
with the possibility to control in situ
the pair interaction
via external fields.\\
Above a periodic potential,
microscopic particles
can be arranged into colloidal
molecular crystals (CMCs)~\cite{Bru02},
i.e. lattices of
doublets, triplets or larger clusters
characterized by
internal rotational degrees of freedom~\cite{Rei02}.
While CMCs are excellent models
to study geometric frustration
effects due to competing orientational order
and lattice constrains~\cite{Agra04,Rei05,Vil05,Sha08},
the focus of these experiments
has been placed mainly on the melting scenario of trimer
systems on a triangular lattice~\cite{Bru02}.
On this lattice, trimers can
be arranged only
in one of two orientational states,
while dimers present a richer
phase behaviour due to the
larger
number of possible
configurations between
pairs~\cite{Vil05}.
The lattice covering by dimer
particles is also
a fascinating problem in statistical mechanics~\cite{Bax82}
which has been recently
the subject of renewed theoretical interest~\cite{Ale05,San06},
in addition
of being present in different processes like
melting~\cite{Sha08},
self-assembly~\cite{Woj91} and molecular adsorption
on crystalline surface~\cite{Ren82}.\\
This letter investigates the colloidal ordering
and dynamics of interacting
microscopic dimers self-assembled
above a honeycomb magnetic lattice.
Each dimer is composed
by a pair of paramagnetic colloids
confined
in a triangular shaped magnetic minimum.
An external precessing
field set the dimers
into rotational motion,
annealing the lattice to a minimum energy state.
Depending on the field
parameters,
the resulting dimer arrangement
can be mapped
to a long range striped phase or to a random fully packed loop (FPL)
state.
On a honeycomb lattice a FPL configuration can be constructed by considering a series 
of arrows which joins the lattice vertices. Each vertex generates one arrow which 
ends into one of the three nearest vertices. All arrows have the same length and 
form a series of closed and self-avoiding loops.
These loops are not allowed to have free
ends, a condition which strongly limit
the number of ways the
arrows can be placed.
FPL models have been used
to explain a broad class of
phenomena
in magnetism, optics an polymer physics~\cite{Blo94,Dup98,Hol08,Nah11,Jau11},
but physical realization are rather scarce.
Dimers on a hexagonal lattice can be effectively arranged
in such a way to produce
a FPL configuration~\cite{Jac09}.
More recently, FPLs have been predicted to appear
for isotropically interacting
colloids arranged above
an honeycomb lattice of triangular
shaped optical traps~\cite{Gia13}.
Here this idealized state is experimentally
reported using magnetic dimers
interacting via dipolar
forces and arranged above a magnetic
lattice.\\
The dimers are composed by pairs of
monodisperse paramagnetic colloids
having diameter $d=2.8 \,{\rm \mu m}$
and magnetic volume susceptibility $\chi \sim 0.4$
(Dynabeads M-270, Dynal).
Due to the doping
with superparamagnetic iron oxide grains,
these particles acquire a dipole moment $\bm{m}=(\pi d^3/6) \chi \bm{H}$,
\begin{figure}[t]
\begin{center}
\includegraphics[width=\columnwidth,keepaspectratio]{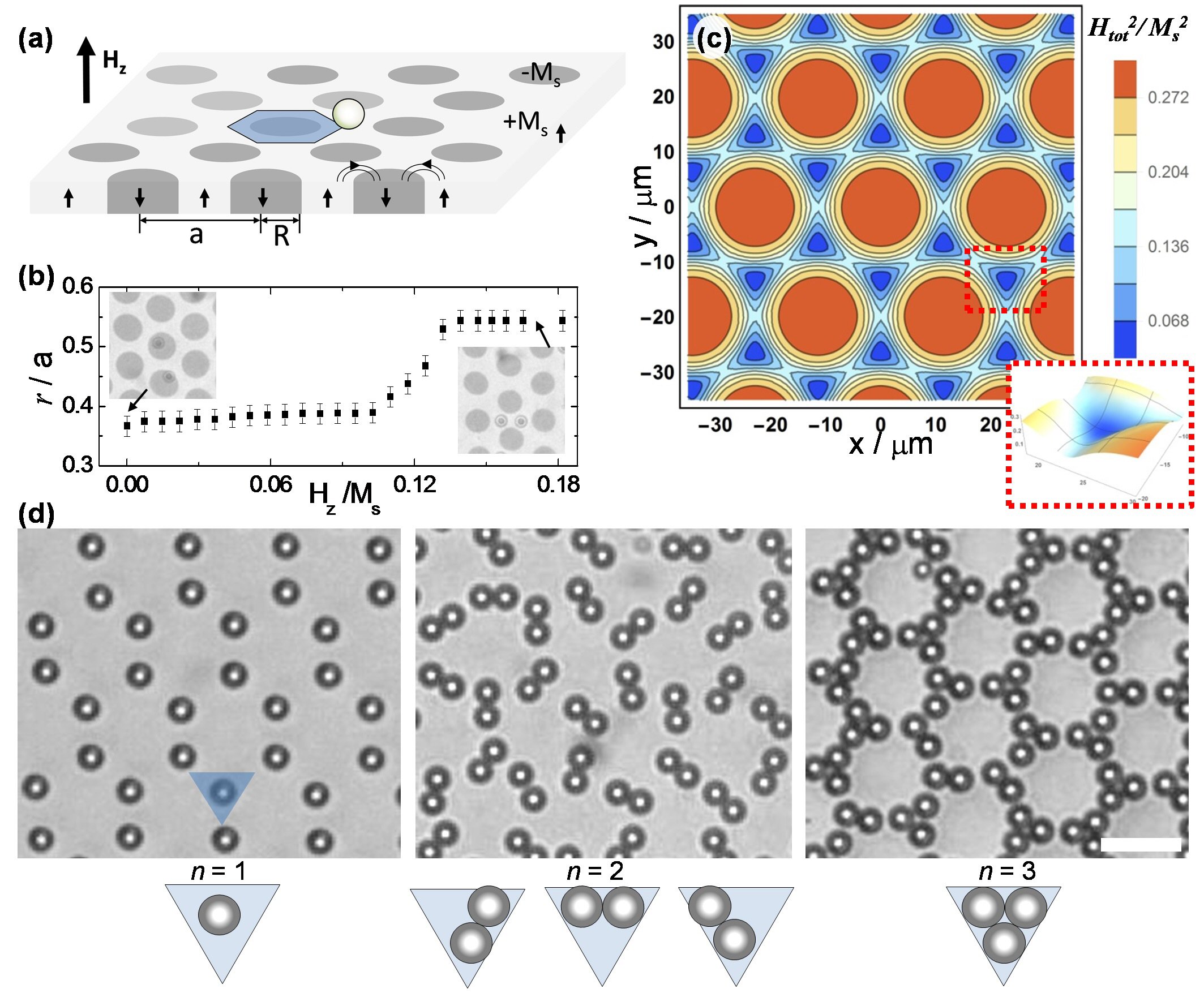}
\caption{(color online)
(a) Schematic of the FGF film
with a magnetic bubble lattice
subjected to
an external field $\bm{H}_z$. One
Wigner-Seitz unit cell is shaded in blue
with one paramagnetic colloid.
For $H_z=0$, the lattice constant is $a=11.6 \,{\rm \mu m}$
and the radius $R=4.2 \,{\rm \mu m}$.
(b) Normalized distance $r/a$ of one
paramagnetic colloid from the center of
a magnetic bubble
versus $H_z$.
(c) Normalized energy landscape
for an FGF under a static field $H_z=0.17  \,{\rm M_s}$.
Inset shows a 3D view of one
up-triangular minimum.
(d) Snapshots of a small
section ($48 \times 42 \,{\rm \mu m^2}$)
of a magnetic bubble lattice
filled with $n=1,2,3$
particles per pinning site (scale bar is $10 \,{\rm \mu m}$).
Schematics at the bottom show
the corresponding
configurations
in a up-triangular minimum.}
\label{figure1}
\end{center}
\end{figure}
when subjected to an external field $\bm{H}$.
The particles are dispersed
in deionized water and
deposited above
a uniaxial ferrite garnet film (FGF)
having thickness $t\sim 4 \mu m$
and saturation magnetization $M_s=1.7 \cdot 10^4 \,{\rm A/m}$~\cite{EPAPS}.
The FGF displays a triangular lattice
of magnetic "bubbles",
i.e. cylindrical ferromagnetic
domains
uniformly magnetized and immersed in an opposite
magnetized film, Fig.1(a).
\begin{figure}[b]
\begin{center}
\includegraphics[width=\columnwidth,keepaspectratio]{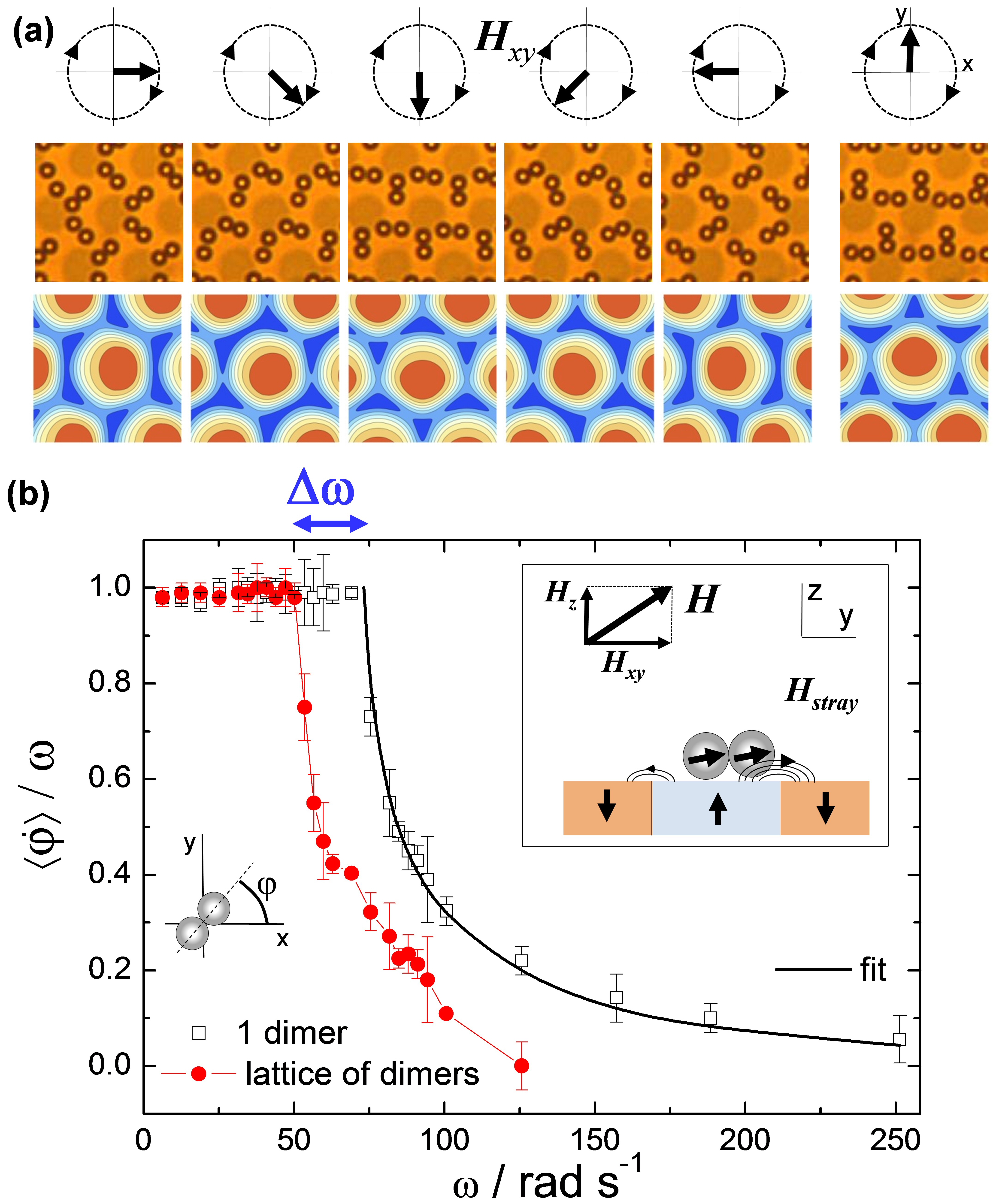}
\caption{(color online)
(a) Polarization microscopy images
showing the locations of the dimers
and of the magnetic bubbles
with directions of the annealing field
($H_{xy}=0.07 \,{\rm M_s}$, $\omega=6.3\,{\rm rad\, s^{-1}}$,
MovieS1).
Images at the bottom show the corresponding
energy landscape deformed by the field,
maxima are colored in red, minima in blue.
(b) Average rotational speed $\langle \dot{\varphi} \rangle$
versus angular frequency $\omega$
for a single dimer, empty squares, and for a
lattice of interacting dimers, filled circles
($H_{xy}=0.08\,{\rm M_s}$).
From the fit (Eq. in the text)
the critical frequency is $\omega^s_c=73.2 \, {\rm rad s^{-1}}$.
Inset:
schematic showing the side view
of a dimer above the FGF.}
\label{figure2}
\end{center}
\end{figure}
The size of the bubble domains
can be easily controlled by
a magnetic field
applied perpendicular to the FGF,
$\bm{H}_z= H_z \bm{e}_z$,
and for amplitudes $H_z\lesssim 0.3 \,{\rm M_s}$
the bubble radius varies linearly with $H_z$
(Fig.S1 in~\cite{EPAPS}).
Once above the FGF surface,
the particles pin at the Bloch wall,
which are located at the
perimeter of the magnetic bubbles.
However for a perpendicular field
$H_z>0.14 \,{\rm M_s}$,
the equilibrium position
of the colloids is shifted
in the interstitial region,
i.e. at the vertices
of the Wigner-Seitz unit cell
around the bubbles, Fig.1(b).
Under these conditions, the location of the
magnetic minima can be visualized by calculating the
energy
of a paramagnetic colloid, $U_m\sim H_{tot}^2$
subjected to the global field $\bm{H}_{tot}=\bm{H}+\bm{H}_{stray}$,
being $\bm{H}$ the applied field and $\bm{H}_{stray}$ the stray field
of the FGF~\cite{EPAPS}.
As shown in Fig.1(c),
the energy landscape
displays  an
honeycomb
lattice of
triangular
shaped minima having alternating orientation.
From the small inset in Fig.1(c), see also
Fig.S5 in~\cite{EPAPS},
it follows that these
minima
have one deep central well
and three higher
wells
at the edges of the triangle.
This feature explains the particle location
at different filling ratio, as shown in Fig.1(d).
With one particle per pinning site
the colloids replicate the honeycomb
lattice of magnetic minima.
Dimers are formed
with two particles,
and have three
energetically equivalent states,
since excluded volume between these
particles impede
to occupy the central well.
In contrast, for trimers this energetic degeneracy
disappears since each particle
can sit close to one of the three
corners of the triangle.
Here we focus on the
dimer system,
which is characterized by frustrated
interactions and a
degenerate ground state.\\
Before each experiment,
we prepare
an initial disordered
configuration of dimers
having a random
distribution of the
three orientations.
The annealing procedures used
to generate this configuration and
later to order the system into
different phases
are both obtained
by superimposing to
$\bm{H}_z$
a rotating
in-plane field,
$\bm{H}_{xy}=H_{xy}[\cos{(\omega t)}\bm{e}_x+\sin{(\omega t)}\bm{e}_y]$.
The resulting applied field, $\bm{H}=\bm{H}_{xy}+\bm{H}_z$ performs
a conical precession around the $z$ axis
with angular frequency $\omega$.
In the preannealing process,
we use higher amplitudes ($H_z=0.25 \,{\rm M_s}$, $H_{xy}=0.20 \,{\rm M_s}$)
such that the modulated landscape
forces exchange of particles between nearest
interstitial and consequently randomizes
the dimer orientations.
After, the
static field is decreased
to $H_z=0.17 \,{\rm M_s}$
and $\bm{H}_{xy}$
is reduced to zero
at a rate of $0.2\,{\rm M_s/s}$.
Once prepared, the lattice of
dimers with disordered orientation
is rather stable with negligible
spontaneous rotations of the dimers
due to thermal fluctuations~\cite{note1}.\\
In order to anneal the lattice into an ordered phase,
a rotating field with amplitude
$H_{xy}\in[0.06, 0.14] {\rm M_s}$ is used,
such that it forces the rotational
motion of the dimers but did not
produced exchange of particles between
consecutive interstitial regions.
Fig.2(a) and MovieS1 in~\cite{EPAPS} show
the effect of a rotating
field with $H_{xy}=0.07 M_s$ on the dimer
orientation.
The bottom row
in Fig.2(a) illustrates how the
energy landscape is
altered by the in-plane field during one cycle. The
applied field modifies the stray field 
of the FGF deforming the triangular minima 
such that they can
accommodate now
a dimer only in one orientation.
The resulting phase is
a long range nematic order 
characterized by
dimers with alternating orientations.
Fig.2(c) shows the average rotational speed
$\langle \dot{\varphi} \rangle$
as a function
of $\omega$
for
a single ($s$) dimer
and of a lattice ($l$)
of dimers due to this annealing
field.
The dynamics for isolated dimer
can be well described as a standard de-synchronization
process in a dissipative medium:
below a critical frequency $\omega^s_c=73.2\,{\rm rad\, s^{-1}}$ the
dimer follows the
rotating field with a constant phase-lag
angle, while
for $\omega>\omega^s_c$ there is an asynchronous regime
where $\langle \dot{\varphi} \rangle$
decreases as the $\omega$ increases.
Neglecting thermal fluctuations, one can fit
well the experimental data using the solution of the deterministic Adler equation~\cite{Adl46},
$\langle \dot{\varphi} \rangle /\omega=1-(1-(\omega^s_c/\omega)^2)^{\frac{1}{2}}$.
In contrast,
at parity of field parameters
the lattice of dimers de-synchronize
earlier, with a critical frequency $\omega^l_c=49.9\,{\rm rad\, s^{-1}}<\omega^s_c$,
and displays a faster decay of $\langle \dot{\varphi} \rangle$
in the asynchronous regime.
When increasing $H_{xy}$
the particle induced moment also
increases since
$\bm{m}\sim (\bm{H}+\bm{H}_{stray})$,
and thus dipolar interactions
between close dimers
become important,
competing with the orientation imposed by the substrate.
In particular,
for in-plane dipoles 
these interactions 
favour alignment
between nearest dimers,
in contrast to the 
ordering induced by the 
rotating landscape where
dimers assemble perpendicular to each other.
Indeed, we find that
the separation frequency,
$\Delta\omega=\omega^s_c-\omega^l_c$
between the two critical frequencies
increases by rising $H_{xy}$, Fig.S2 in~\cite{EPAPS}.
As shown in the small inset in Fig.2(b),
the induced moments in the
particles are mainly oriented
in the plane of the FGF due to
the configuration
of the stray field.
On the surface of the FGF, the $\bm{H}_{xy}$ field
strengthens (or weakens) the stray field above the interstitial region
depending whether the magnetic field lines
are parallel (or antiparallel)
to the applied field.\\
Next we explore the stationary phases
which emerge when a disordered lattice of dimers
is subjected to an annealing by varying 
the amplitude and frequency of $\bm{H}_{xy}$.
In order to characterize the dimer arrangement,
we use the arrow representation as originally
introduced by Elser and Zeng~\cite{Els93}
for the spin-$\frac{1}{2}$ kagome antiferromagnet,
and shown in Fig.3(a).
\begin{figure}[t]
\begin{center}
\includegraphics[width=\columnwidth,keepaspectratio]{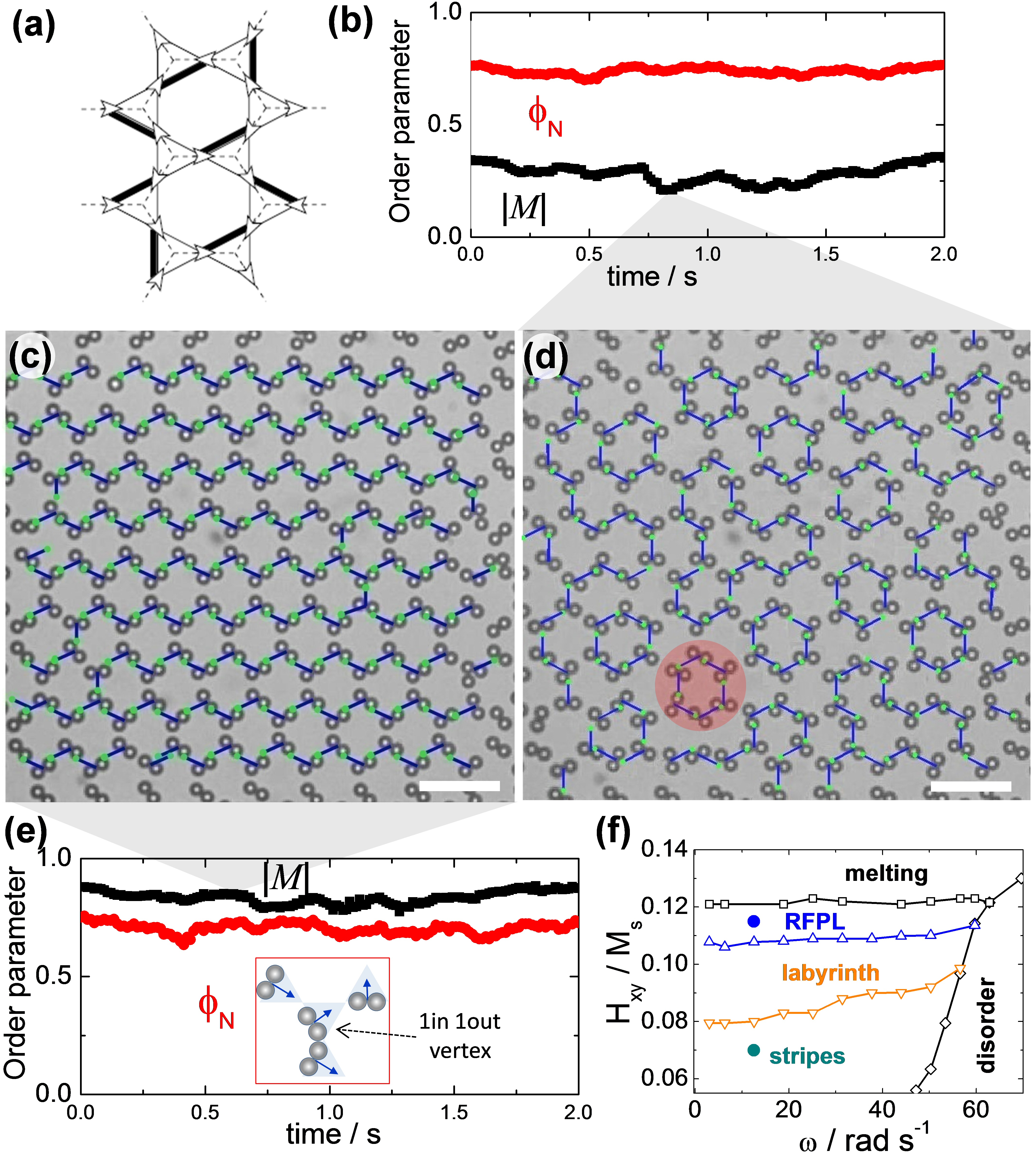}
\caption{(color online) (a)
Arrow representation of dimer covering
a kagome lattice,
image reproduced from Ref.~\cite{Mis02}.
(b) Order parameters $|\bm{M}|$ and $\phi_N$
{\it vs} time for Fig.3(d).
(c) Stripe phase of a lattice
of dimers ($H_{xy}=0.07\,{\rm M_s}$, $\omega=3.1\,{\rm rad\, s^{-1}}$).
To each dimer is assigned a blue
arrow with a green dot as head
following the
arrow representation.
(d) Random FPL phase ($H_{xy}=0.11\,{\rm M_s}$, $\omega=3.1\,{\rm rad\, s^{-1}}$),
one small loop is highlighted  in red.
(e) Order parameters $|\bm{M}|$ and $\phi_N$
{\it vs} time for Fig.3(c).
Inset: schematic showing the definition of $\phi_N$,
as the fraction of vertices having one incoming and one outgoing arrow.
(f) Diagram
in the ($\omega$,$H_{xy}$)
plane
illustrating the various colloidal
phases. Filled points indicate locations of
Fig.3(c) and (d)}
\label{figure3}
\end{center}
\end{figure}
Since each
dimer sit on
one of the three sides
of a triangular minimum,
to each dimer can be uniquely associated
an arrow pointing
from the dimer center to the free corner of the triangle.
Each triangle has three nearest neighbors,
one outgoing arrow
and from $0$ to $3$ incoming ones.
In the latter case, two
arrows from adjacent triangles
can superimpose and create a defect.
FPLs occur when
there are no such defects and the arrows form
closed loops which visit each lattice vertex only once.
Two representative images
showing a long range striped
phase and a random FPL state
obtained with this mapping 
are shown in Figs.3(c,d),
where to guide the eyes
the arrows have a green dot as head.
The first state, already described in Fig.2(a),
is characterized by
parallel
stripes of arrows
with mean director
given by the orientation of the
applied field, which breaks the symmetry of the
underlying potential.
Sliding symmetries
characterized by parallel stripes
having opposite directions
as observed in~\cite{Gia13}
are not possible here.
A typical labyrinth
and a disordered phase are showed in~\cite{EPAPS}.
To distinguish between the various phases,
we measure two order parameters:
a N\'eel type parameter $|\bm{M}|$~\cite{note2},
and the fraction $\phi_N$
of vertices in the lattice having exactly one
incoming and
one outgoing arrow,
as defined in the schematic of Fig.3(e).
The nematic ordering
is characterized by high values of both order
parameters.
Random
FPLs have lower
value of $|\bm{M}|$
since the stripes
break into smaller
loops, but conserve the high
fraction $\phi_N$, Fig.3(b).
In contrast, disordered states are characterized by
low values of both parameters.
Moreover, from the diagram
shown in Fig.3(f), emerges that
the transition between these
phases depends weakly on $\omega$
and mainly on $H_{xy}$
which controls the interaction strength.
The stripe ordering is observed for $H_{xy} \lesssim 0.8 \,{\rm M_s}$.
Increasing $H_{xy}$,
the dimers interacts
strongly and the stripes start
to break up forming
an intermediate labyrinth like pattern (Fig.S3 in~\cite{EPAPS}).
The bending of the stripes at high value
of $H_{xy}$
can be understood by
considering the effect of the time-averaged dipolar interactions~\cite{note3}.
In absence of FGF,
these interactions are attractive
and force a chain of
particles like pairs of dimers,
to aggregate into a compact cluster~\cite{Cas10}.
The presence of the honeycomb lattice prevents
the formation of these clusters,
but
the stripes can more easily break
due to the loss of synchronization
of some composing dimer.
For $H_{xy}\sim 0.11 \,{\rm M_s}$,
the strings of arrows
completely break into
smaller FPLs
randomly distributed
above the film.
In particular,
in many FPL state found,
we observe a large fraction of elementary
loops formed
by six touching arrows which can
have both
sense of rotations, Fig.3(d).
For field larger than $H_{xy}\gtrsim 0.12 \,{\rm M_s}$,
\begin{figure}[t]
\begin{center}
\includegraphics[width=\columnwidth,keepaspectratio]{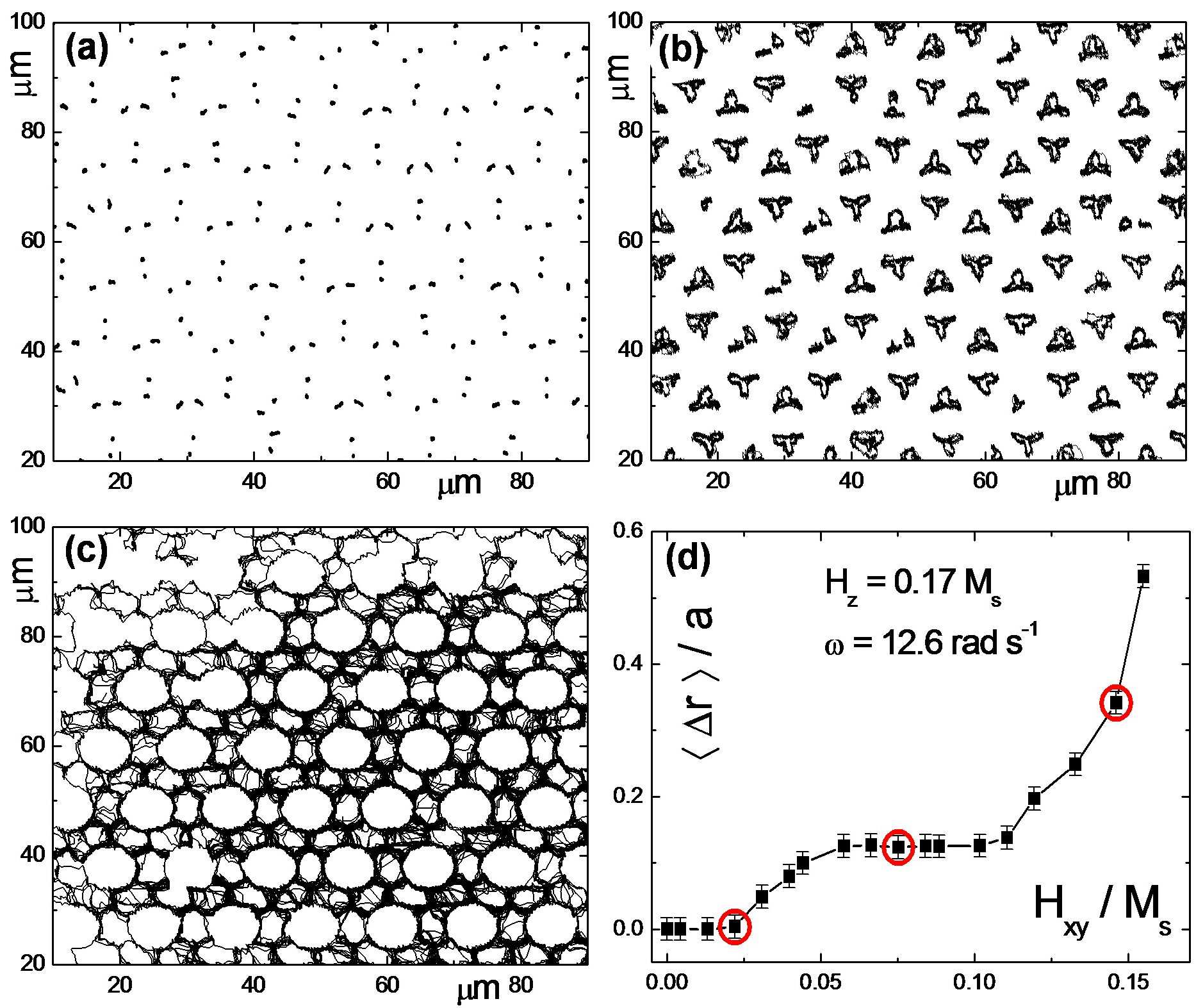}
\caption{(color online)(a-c) Colloidal
trajectories (lines) for dimers subjected to
an external
precessing field
with angular frequency $\omega = 12.6 \,{\rm rad\, s^{-1}}$,
and amplitudes $H_z=0.17 \,{\rm M_s}$ and $H_{xy}=0.02 \,{\rm M_s}$
for (a) [MovieS2 in~\cite{EPAPS}], $H_{xy}=0.07 \,{\rm M_s}$ for (b) [MovieS3 in~\cite{EPAPS}]; 
$H_{xy}=0.14 \,{\rm M_s}$ for (c) [MovieS4 in~\cite{EPAPS}].
(d) Average displacement $\langle \Delta r \rangle$
versus amplitude of the in-plane rotating field $H_{xy}$.
Grey (red) circles denotes the location of Figs.4(a-c).}
\label{figure4}
\end{center}
\end{figure}
these interactions are so strong that they induce complete 
melting of the lattice.\\
Finally, we explored the melting process
in the dimer lattice which can be induced by either
decreasing $H_z$ or,
as commented previously, by
further increasing $H_{xy}$ 
such that
dipolar interactions
completely dominate.
The latter case is illustrated in Figs.4(a-c),
and it features a two stage melting
transition.
For $H_{xy}<0.03 \,{\rm M_s}$, the applied field is unable
to rotate completely the dimers,
and the system
remain frozen in the initial ordered state, Fig.4(a).
Increasing $H_{x,y}$ induces
a transition towards a partially ordered
state,
when the dimers perform localized rotational motion,
but the system keeps its
positional
order, Fig.4(b).
Finally, for $H_{xy} \gtrsim 0.12 \,{\rm M_s}$
a second disorder transition
occurs when the dimers
break up and reform
exchanging particles as the field
is rotating, MovieS4 in~\cite{EPAPS}.
The system
forms a
liquid phase where strong
attractive dipolar forces
favour exchange of particles
between nearest interstitial when two
dimers rotate close to each other.
To quantify this two-step
transition,
the globally averaged particle displacement,
$\langle \Delta r \rangle$
is measured for each applied field.
We find that both transitions are rather
smooth, second
order-like and characterized by
the presence of a finite step
of the order parameter in the intermediate phase,
Fig.4(d).
A similar melting scenario with finite steps in $\langle \Delta r \rangle$
has been previously predicted via numerical
simulations~\cite{Rei05}.
However melting here is induced by increasing
the dipolar coupling between the
particles, rather than decreasing the substrate strength
in favour of thermal fluctuations~\cite{Bru02,Rei02}.\\
In summary, we realize a frustrated
colloidal molecular crystal
composed by
self-assembled microscopic dimers
interacting above a magnetic honeycomb lattice.
The system reveals a rich phase behaviour
when dimer-dimer interactions compete with
substrate strength.
These interactions can be tuned in situ
via application of a rotating field.
The dimer covering
of periodic lattices
can be mapped
to Ising systems~\cite{Fis61,Kas63},
or can be used as simplified
model for the
adsorption of diatomic molecules onto a surface,
like $N_2$ on graphite~\cite{Mou82},
lattice gas systems~\cite{Loh13}
and tiling problems~\cite{Mou82}.
Yet the transport properties of bound dimers
on
a periodic lattice~\cite{Geh09},
such as DNA linked colloidal doublets~\cite{Tie08},
is also an interesting
future avenue which can be explored
with the presented system.\\
I acknowledge T. M. Fischer for stimulating discussions
and Tom H. Johansen for the FGF. This work
was supported by the European
Research Council via Project No. 335040 and by the
"Ramon y Cajal" program (No. RYC-2011-07605).
\bibliography{biblio}

\begin{thebibliography}{55}
\bibitem [{not(){\natexlab{a}}}]{note1}%
  \BibitemOpen
  \href@noop {} {} ({\natexlab{a}}),\ \bibinfo {note} {on a striped FGF the
  effective translational diffusion coefficient was measured as $D=3.2 \cdot
  10^{-4} \,{\rm \mu m^2 s^{-1}}$~\cite{Tie10}.}\BibitemShut {Stop}%
\bibitem [{not(){\natexlab{b}}}]{note2}%
  \BibitemOpen
  \href@noop {} {} ({\natexlab{b}}),\ \bibinfo {note} {in analogy with
  Ref.~\cite{Gia13}, form the position of the particles in the dimers and from
  the corresponding position of the nereast magnetic bubble, we calculate the
  N\'eel parameter as $\bm{M}=2/(\sqrt{3}N)\sum_i(-1)^i\hat{\bm{e}}_i$, being
  $\hat{\bm{e}}_1\equiv(0,1)$, and
  $\hat{\bm{e}}_{2,3}\equiv(\pm\sqrt{3}/2,-1/2)$ the three lattice vectors,
  $(-1)^i$ is $+1$ ($-1$) for an up (down) triangle.}\BibitemShut {Stop}%
\bibitem [{not(){\natexlab{c}}}]{note3}%
  \BibitemOpen
  \href@noop {} {} ({\natexlab{c}}),\ \bibinfo {note} {the normalized pair
  potential between two induced dipoles $\bm{m}_i$, $\bm{m}_j$ separated by a
  distance $\bm{r}_{ij}=\bm{r}_i-\bm{r}_j$ can be written as,
  $U_{m}=(\mu_m/4\pi)[(\bm{m}_i \cdot \bm{m}_j)/r^3_{ij}-3(\bm{m}_i \cdot
  \bm{r}_{ij})(\bm{m}_j \cdot \bm{r}_{ij})/r^5_{ij}]$, being $\mu_m$ the
  magnetic susceptibility of the medium. Performing a time average of the
  dipolar energy between two colloids when subjected to a rotating field
  $\bm{H}_{xy}$ in the $(x,y)$ plane gives an effective attractive potential in
  this plane, $\langle U_{m} \rangle =(-\mu_m m^2/8\pi (x+y)^3)$.}\BibitemShut
  {Stop}%
\end{thebibliography}
\end{document}